\documentclass[twocolumn,prc,floatfix,showpacs,preprintnumbers,amsmath,amssymb,showkeys]{revtex4}

\usepackage{graphicx}
\usepackage{dcolumn}
\usepackage{bm}

\begin{document}



\title{Analysis of bulk and surface contributions in the neutron skin
of nuclei}

\author{M. Warda\textsuperscript{1,2}}
 \email{warda@kft.umcs.lublin.pl}
\author{X. Vi\~nas\textsuperscript{1}}
 \email{xavier@ecm.ub.es}
\author{X. Roca-Maza\textsuperscript{1}}
 \email{roca@ecm.ub.es}
\author{M. Centelles\textsuperscript{1}}
 \email{mariocentelles@ub.edu}

\affiliation{\textsuperscript{1}Departament d'Estructura i Constituents de la Mat\`eria
and Institut de Ci\`encies del Cosmos,
Facultat de F\'{\i}sica, Universitat de Barcelona,
Diagonal {\sl 647}, {\sl 08028} Barcelona, Spain\\
\textsuperscript{2}Katedra Fizyki Teoretycznej, Uniwersytet Marii Curie--Sk\l odowskiej,
        ul. Radziszewskiego 10, 20-031 Lublin, Poland}

\date{\today}

\begin{abstract}
The neutron skin thickness of nuclei is a sensitive probe of the
nuclear symmetry energy and has multiple implications for nuclear and
astrophysical studies. However, precision measurements of this
observable are difficult. The analysis of the experimental data may
imply some assumptions about the bulk or surface nature of the
formation of the neutron skin. Here we study the bulk or surface
character of neutron skins of nuclei following from calculations with
Gogny, Skyrme, and covariant nuclear mean-field interactions. These
interactions are successful in describing nuclear charge radii and
binding energies but predict different values for neutron skins. We
perform the study by fitting two-parameter Fermi distributions to the
calculated self-consistent neutron and proton densities. We note that
the equivalent sharp radius is a more suitable reference quantity than
the half-density radius parameter of the Fermi distributions to discern
between the bulk and surface contributions in neutron skins. We
present calculations for nuclei in the stability valley and for the
isotopic chains of Sn and Pb.

\end{abstract}

\maketitle


\section{Introduction}


The description of the sizes and shapes of atomic nuclei is among the
oldest problems in nuclear physics. The rms radius
of the charge distribution in nuclei surely is the most prominent
example of this kind of observable. Owing to the high degree of
accuracy achieved by the elastic electron-nucleus and muon-nucleus
scattering experiments, the nuclear charge radius is nowadays known
with uncertainties that are for many nuclei smaller than 1\%
\cite{fricke95,ang04}. In contrast, our knowledge about the neutron
distribution and its rms radius in nuclei is not so precise. Actually,
accurate determinations of the rms radius of the neutron density are
still lacking. This implies that the so-called neutron skin thickness,
generally defined as the neutron-proton rms radius difference in the
atomic nucleus,
\begin{equation}
\label{skin}
\Delta R_{np}= \langle r^2 \rangle_n^{1/2}
- \langle r^2 \rangle_p^{1/2} ,
\end{equation}
is not  precisely known either.

The neutron skin thickness observable, Eq. (\ref{skin}), is a very sensitive
probe of the pressure difference that exists between neutrons and
protons in the atomic nucleus. As such, $\Delta R_{np}$ is intimately
correlated with the density dependence of the nuclear symmetry energy
and with the equation of state of pure neutron matter
\cite{bro00,typ01,cen02,fur02,dan03,die03,bal04,ava07,cen09,cen09b}.
Owing to this fact, the accurate calibration of $\Delta R_{np}$ is a
problem of significant implications for studies that embrace diverse
facets of both nuclear physics and nuclear astrophysics (see for
example Refs.\
\cite{hor01,sil05,ste05a,lat07,kli07,li08,piek09,xu09,sun09,vid09,car10}). It should
be mentioned that Eq.\ (\ref{skin}) is not the only useful
prescription to characterize the different spatial extension of the
neutron and proton densities in a nucleus. The
neutron-proton radius difference has also been  computed using
Helm radii instead of rms radii for the discussion of nuclear halos in
the literature \cite{miz00,sch08}. Nevertheless, in the present work
we use the conventional and more frequent
definition (\ref{skin}) for $\Delta R_{np}$, Eq. (\ref{skin}).

Because neutrons are uncharged particles, the measurement of their
spatial distribution in the nucleus is more difficult than for the
positively charged protons (see, e.g., Ref. \cite{bat89}). 
The experimental access to neutron
densities and neutron rms radii usually involves strongly interacting
hadronic probes, for example, in the case of experiments that
perform proton-nucleus elastic scattering
\cite{ray78,ray85,sta94,kar02,cla03}, $\alpha$-particle elastic scattering \cite{gil84}, and techniques based on
the inelastic scattering excitation of the giant dipole and
spin-dipole resonances \cite{kra99,kra04}. The neutron skin thickness
of nuclei  has also been investigated through radiochemical and x-ray
techniques in antiprotonic
atoms~\cite{trz01,jas04,lub94,lub98,sch99,klo07},
taking advantage of the fact that the nuclear periphery is very
sensitive to antiprotons in the normally electronic shell. Though the
scope of our analysis of neutron skins is of theoretical, in this
article we refer to some extent to the experimental
investigations in antiprotonic atoms to set the stage for our
calculations.

A few years ago, Trzci{\'n}ska {\it et al.}\ \cite{trz01,jas04} extracted
the neutron skin thickness of a large set of nuclei in experiments
with antiprotons conducted at the former LEAR facility of CERN\@. The
measurements were made for 26 stable isotopes distributed across the
mass table, from the light and symmetric nucleus $^{40}$Ca to the heavy
and asymmetric nucleus $^{238}$U. Because of the fact that the antiproton-nucleon
interaction is very strong, antiprotons are able to interact with the
atomic nucleus at distances where the nuclear density is much smaller
than its central value. Slow enough antiprotons can form a hydrogen-like 
atom with the 
nucleus. When the antiproton annihilates with a nucleon
producing pions that may miss the nucleus, it leaves a residue that is one
neutron or proton fewer. From the analysis of these yields, information
about the neutron distribution in the nucleus can be obtained
\cite{trz01,jas04,lub94,lub98,sch99}. A second experimental method
measures antiprotonic x-rays from where the atomic level widths and
shifts owing to the strong interaction are determined
\cite{trz01,jas04,klo07}. By combining the results obtained with these
two experimental techniques, the neutron-proton rms difference $\Delta
R_{np}$ can be deduced provided that the charge density of the nucleus
is known~\cite{trz01,jas04,klo07}.

The extraction of $\Delta R_{np}$ values from antiprotonic atoms
assumes nucleon densities in the form of two-parameter Fermi (2pF)
distributions. The procedure involves interpreting whether the
difference between the peripheral neutron and proton densities arises
from an increase of the mean location of the surface of the neutron
density (i.e., from an increase of the bulk radius of neutrons) or,
rather, from an increase of the surface diffuseness of the neutron
density. This question is also instrumental in studies of properties
of nuclei by parity-violating electron scattering \cite{don09}. The
radiochemical data in antiprotonic atoms were shown to be in favor of
interpreting $\Delta R_{np}$ as an increase of the neutron surface
diffuseness \cite{trz01,jas04} but with the caveat that some room
existed within assigned errors for an intermediate situation
\cite{trz01}. Indeed, a comparison with the droplet
model~\cite{swi05} suggests that the neutron skin sizes of the
antiprotonic measurements can be described similarly well in the
droplet model theory by a difference in the diffuseness as by a
difference in the bulk radius of the neutron and proton densities.

In the present work we theoretically investigate the bulk and surface
components in the neutron skin thickness of nuclei by parametrizing
self-consistently calculated nucleon densities by 2pF distributions.
The 2pF form is also common in experimental analyses. Our calculations
are performed with some representative effective nuclear forces. These are
the finite-range Gogny D1S interaction and the zero-range Skyrme SLy4
force from the nonrelativistic framework, and the NL3 and FSUGold
parameter sets from the relativistic mean-field (RMF) framework. We
discuss the predictions that these mean-field models make for the
decomposition of the neutron skin thickness in bulk and surface
contributions for the set of stable nuclei that were analyzed in the
experiments in antiprotonic atoms. We also study the theoretical
predictions along the isotopic chains of Sn and Pb and the variation
of the results as one moves from the proton to the neutron drip line in
these isotopic chains.

The structure of this article is the following. In the Sec. II we
present a brief summary of the experimental methodology and results in
antiprotonic atoms to highlight some interesting aspects for our
study. In Sec. III we discuss the common definitions of
nuclear radii and characterize the bulk and surface
contributions in the neutron skin thickness of nuclei. In Sec.~IV 
we analyze the theoretical mean-field results in stable isotopes and
the predictions for nuclei across the Sn and Pb isotopic chains. We
present the summary and conclusions in Sec.~V\@. Some relations for
2pF functions are collected in the Appendix.


\section{Some aspects of the methods and results in
antiprotonic atoms}


The radiochemical study of antiprotonic atoms
\cite{trz01,jas04,lub94,lub98,sch99,wyc07,fri05,fri09} 
consists of the analysis of the
nuclei with a mass number one unit smaller than the target mass number
$A_t$ where the antiproton annihilation takes place. These products
that have one less neutron or proton are short lived and their decay is
followed by emission of $\gamma$ rays. Standard nuclear spectroscopy
methods allow one to determine the absolute number of these residual
nuclei with mass number $A_t-1$. The probability of producing a cold
product with mass number $A_t-1$ is calculated by using the
antiproton-nucleus optical potential fitted to reproduce the x-ray
experimental data (atomic level shifts and level widths)
\cite{trz01,klo07,bat95}. The probability distribution for obtaining a
nucleus with mass number $A_t-1$ has a maximum located about 2--3~fm
outside the half-density radius $R_{1/2}$ of the target nucleus.

To compare the experimental data for any target it is convenient
to introduce the peripheral halo factor~\cite{bug73}:
\begin{equation}  
\label{fhalo} 
f_{\rm halo}^{\rm expt}= 
\left[ \frac{N(\bar p \, n)}{N(\bar p \, p)} \, \frac{Z_t}{N_t} 
\right] 
\bigg/
\left[ \frac{{\rm Im}(\delta_{\bar p n})}{{\rm Im}(\delta_{\bar p p})}
\right] .
\end{equation} 
The first term in brackets on the right-hand side of this equation, gives the
ratio of the $\bar p$ annihilations on peripheral neutrons to the
$\bar p$ annihilations on peripheral protons, normalized with the
$Z_t/N_t$ value of the target nucleus. The quantities $\delta_{\bar p
n}$ and $\delta_{\bar p p}$ are the $\bar p$-$n$ and $\bar p$-$p$
scattering amplitudes, respectively, and the factor $F= {\rm Im}
(\delta_{\bar p n})/{\rm Im}(\delta_{\bar p p})$ gives the ratio of
the $\bar p$ annihilation probabilities on a neutron and on a proton.
Consequently, halo factor (\ref{fhalo}) essentially measures the
change of the ratio of the neutron-to-proton concentration in the
peripheral region with respect to the bulk value represented by the
$N_t/Z_t$ ratio in the target nucleus. A halo factor larger (smaller)
than 1 means an increase of the relative neutron (proton)
concentration in the nuclear periphery. Assuming the ratio $F$ is
known, the measurement of the $N_t-1$ and $Z_t-1$ product yields
allows one to obtain the experimental value of the halo factor. It was
concluded that the comparison of the results from the antiprotonic
x-ray method with the radiochemical method favors a value of $F\approx 1$
\cite{trz01}.

The number of nuclei withone fewer  neutron or  proton  is proportional
to the quantities ${\rm Im}(\delta_{\bar p n})\rho_n$ or ${\rm
Im}(\delta_{\bar p p}) \rho_p$, respectively, integrated over the
region where the annihilation process takes place. Thus,
approximately, one can estimate theoretically the halo factor as
\cite{sch99,bar97} 
\begin{equation} 
\label{fhaloapp}
f_{\rm halo}^{\rm theor}(r) \approx \frac{\rho_n(r)}{\rho_p(r)}
\frac{Z_t}{N_t} \,.
\end{equation} 
This factor  reaches the experimental value $f_{\rm halo}^{\rm expt}$ at a
distance $r\approx R_{1/2}+2.5$~fm beyond the half-density radius of the charge
distribution, where the antiproton annihilation takes place. 
One can use this fact to reproduce the neutron density distribution
from some experimental observables.

The charge distribution in many stable nuclei is known very precisely.
Usually the experimental charge density is given in some analytical
form that is or may be converted to a 2pF
distribution \cite{vri87,has88}. The simple 2pF formula
\begin{equation} 
\label{2pf}
\rho(r) = 
\frac{\rho_0}{1+ \exp{ [(r-C)/a] }} 
\end{equation} 
has only two free parameters with clear physical meaning: on the
one hand, $a$ describes the diffuseness of the surface of the density
profile: on the other hand, the half density or central radius $C$
describes the mean location of this surface (i.e., $C$ is indicative
of the extension of the bulk part of the density distribution). 
The other, more
sophisticated, parametrizations of the density distributions
\cite{vri87,has88} describe better the central part of the nucleus or
modify the surface part with higher-order terms. We do not use them
because the interpretation of the multiple parameters is harder and less
direct. 

To study the differences at the nuclear periphery between the neutron
and proton densities in the antiprotonic atoms experiments, the
authors of Ref.\ \cite{trz01} used 2pF functions. They used the notation
``neutron skin-type'' distribution and  ``neutron halo-type''
distribution to describe the two extreme cases of 2pF shapes having either
$C_n>C_p$ and $a_n=a_p$, or $C_n=C_p$ and $a_n>a_p$, respectively. As
we mentioned in the Introduction, the radiochemical data gave
support to understanding the difference between the neutron and proton
rms radii in the antiprotonic atoms from an increase of $a_n$ rather
than from an increase of $C_n$, compared to the $a_p$ and $C_p$
values. Thus, the neutron halo-type distribution ($C_n-C_p=0$) was
assumed \cite{trz01}. (Note that here ``halo type'' is a useful
notation, but the nuclei in the antiprotonic experiments are stable
isotopes and it is not meant that they have halos like very light or
exotic nuclei, such as $^{11}$Li \cite{miz00,sch08}). It was noted in the same work
\cite{trz01} that some room was left, nevertheless, within the error
bars for intermediate cases with $C_n>C_p$ and $a_n>a_p$.

\begin{figure}
\includegraphics[width=0.95\columnwidth,clip=true]{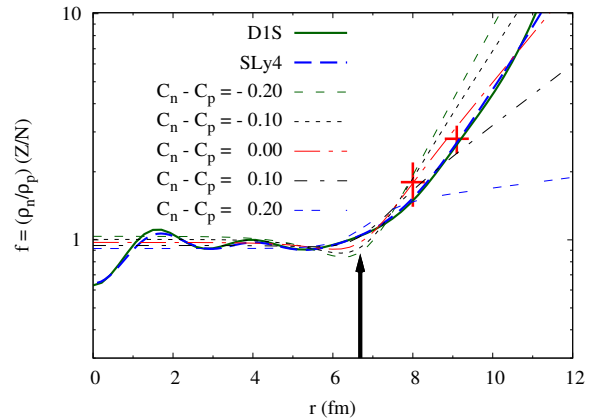}
\caption{\label{FIGURE1} (Color online)
The halo factor in $^{208}$Pb as a function
of the distance from the center of the nucleus, calculated using the
2pF formulas from the experimental charge density \cite{fri95}
assuming a neutron skin thickness $\Delta R_{np}=0.16$ fm and
different values of $C_n-C_p$ (thin lines). The results of the
theoretical predictions with the Gogny D1S and Skyrme SLy4 nuclear
forces are also shown. The values of the halo factor deduced from
experiment \cite{trz01,klo07} are marked by crosses. The value of the
proton half-density radius $C_p$ is indicated by an arrow.}
\end{figure}

The 2pF shape can be applied for the description of charge, proton, or
neutron densities. Indeed, if the charge density is known in the 2pF
form, the corresponding point proton density can be easily found and
it also takes  a 2pF form, with the parameters obtained through the
deconvolution procedure \cite{gar92,pat03}. The more relevant expressions
are given in the Appendix. From these proton
density profiles one can try to deduce the neutron density profiles
with some additional information. For instance, if one knows the
$C_p$ and $a_p$ parameters of the proton density and the value of the
neutron skin thickness $\Delta R_{np}$ defined in Eq.\ (\ref{skin})
(e.g., from independent experimental measurements), a relationship
between the half-density radius $C_n$ and the diffuseness $a_n$ of the
neutron 2pF distribution can be found (see the Appendix). Such a
procedure creates a family of neutron density profiles that depend on
one free parameter, which can be chosen to be either $C_n-C_p$ or
$a_n-a_p$. Now it is possible to check which set of values of $C_n$
and $a_n$ reproduces the experimental halo factor (\ref{fhalo}). This
allows one to determine the 2pF neutron distribution. Next we apply
this idea to the $^{208}$Pb nucleus as an illustrative test case.

In Fig.~\ref{FIGURE1} we display the halo factor, Eq. (\ref{fhaloapp}), for
the $^{208}$Pb nucleus computed with 2pF density distributions. 
The 2pF neutron densities are obtained from the experimental charge
density taken from Ref.\ \cite{fri95}, which we convert to a
2pF proton density using the formulas given in the Appendix. We 
consider a family of 2pF neutron densities that differ one from the
other in the half-density radius $C_n$ and diffuseness $a_n$, but they
all have the same value for the neutron skin thickness: $\Delta
R_{np}=0.16$~fm (which corresponds to the value extracted from the
results of the x-ray method in the $^{208}$Pb antiprotonic atom
\cite{klo07}). We allow the difference $C_n-C_p$ to vary in the
range from $-0.20$ to $0.20$~fm. The value of the diffuseness $a_n$
of the 2pF neutron density is then obtained using
Eq.~(\ref{EQUATION_A11}) from the Appendix.

\begin{table}
\caption{\label{TABLE1} The parameters $C$ and $a$ of the 2pF
distributions for $^{208}$Pb obtained from the experimental charge
density \cite{fri95} as described in the text and from mean-field
calculations. All values are given in femtometers.}
\begin{ruledtabular}
\begin{tabular}{lcccc}
& $C_q$ & $a_q$ & $C_n-C_p$ & $a_n-a_p$ \\
\hline
proton 	& 6.704  & 0.438 &&\\
neutron & 6.504  & 0.659 & $-$0.20 & 0.221 \\
 	& 6.604  & 0.614 & $-$0.10 & 0.176 \\
 	& 6.704  & 0.565 &    0.00 & 0.127 \\
	& 6.804  & 0.511 &    0.10 & 0.073  \\
	& 6.904  & 0.449 &    0.20 & 0.011 
\\
\hline
\multicolumn{5}{c}{fit to D1S density profile}\\
\hline
proton 	& 6.645  & 0.467 \\
neutron & 6.686  & 0.548 & 0.041 & 0.081 \\
\hline
\multicolumn{5}{c}{fit to SLy4 density profile}\\
\hline
proton 	& 6.683  & 0.470 \\
neutron & 6.755  & 0.555 & 0.072 & 0.085
\end{tabular}
\end{ruledtabular}
\end{table}

The values of the 2pF parameters in our present calculation are given
in Table~\ref{TABLE1}. Agreement with the results from previous
experiments \cite{trz01,klo07} (indicated by crosses in
Fig.~\ref{FIGURE1}) is found for the parameters of the 2pF
distributions in the range from the values $C_n-C_p=0$ fm and
$a_n-a_p=0.13$~fm to the values $C_n-C_p=0.10$~fm and
$a_n-a_p=0.07$~fm. In the first limit, the neutron skin thickness
$\Delta R_{np}$ is basically due to an enhancement of the diffuseness
of the neutron density, whereas the second limit corresponds to a
mixed configuration where both the neutron diffuseness and the neutron
half-density radius are larger than the values of the proton density.
We have to stress that even small differences between $C_n$ and $C_p$
may have a meaningful influence on the value of the neutron skin
thickness. The discussed results indicate an intermediate character of
the neutron skin thickness in $^{208}$Pb, but with a preference for the
density pattern where $C_n \approx C_p$ and $a_n > a_p$. A more
precise conclusion is difficult because of the large experimental error
bars.

For comparison, in Fig.~\ref{FIGURE1} we also plot the
theoretical halo factor (\ref{fhaloapp}) in $^{208}$Pb calculated directly from
the neutron and proton densities of the Gogny D1S \cite{ber91} and
Skyrme SLy4 \cite{cha98} effective nuclear forces. At the distances
relevant for antiproton annihilation, the halo factor obtained
with these theoretical mean-field densities agrees considerably well
with the experimental data. The numerical values of the half-density
radius $C$ and of the diffuseness parameter $a$ of the D1S and SLy4
equivalent 2pF neutron and proton densities for $^{208}$Pb are
reported in Table~\ref{TABLE1}. For either force, one observes that
the half-density radii $C_n$ and $C_p$ of the 2pF distributions are
different and that the values of $a_n$ and $a_p$ are also different.
Thus, the results of these models correspond to some
intermediate situation between the ``neutron halo-type'' 2pF
distribution (where $C_n=C_p$ and $a_n>a_p$) and the ``neutron
skin-type'' 2pF distribution (where $C_n>C_p$ and $a_n=a_p$) discussed
in Ref.\ \cite{trz01}. In any case, the two forces, D1S and SLy4, show
some preference, as also do the calculations considered earlier, for the
``neutron halo-type'' distribution in $^{208}$Pb, especially in the
case of the D1S force, where $C_n-C_p= 0.04$~fm and $a_n-a_p= 0.08$~fm.

From the discussions of this section it is easily seen that the
characterization of the ``bulk'' or ``surface'' formation of the
neutron skin thickness in nuclei is a nontrivial problem. In the
following, we wish to analyze this question according to the
predictions derived from mean-field models of nuclear structure that
are tested to be successful for charge radii, binding energies, and a
wealth of phenomena in nuclei. First, we need to establish a
prescription to separate bulk and surface contributions in neutron
skins calculated with mean-field nuclear densities.


\section{Discerning bulk from surface in neutron skins}
\label{discerning}


Nuclear radii can be defined in several independent ways. The most
popular formulas and their relations are discussed in the book by
Hasse and Myers \cite{has88} and we recall and analyze them in this
section. One of the simplest ways to describe the size of nuclei is to
define a {\it central radius} $C$ in terms of the integral of the
nuclear density profile $\rho(r)$ (for either neutrons or protons) as
\begin{equation} 
\label{c}
C= \frac{1}{\rho(0)} \int_0^{\infty} \rho(r) dr \,.
\end{equation} 
Another option is the {\it half-density radius} $R_{1/2}$ which is
defined from the local condition
\begin{equation}
\label{r12}
\rho(R_{1/2})=\frac 12 \, \rho(0) \,.
\end{equation}
For density profiles that have a symmetric surface, such as the 2pF
distribution defined in Eq.\ (\ref{2pf}), the central radius $C$ and
the half-density radius $R_{1/2}$ coincide.

The {\it equivalent sharp radius} $R$ is the radius of a uniform sharp
distribution that has a constant density equal to the bulk value of the
actual density and that contains the same number of nucleons as the
considered nucleus:
\begin{equation} 
\label{r}
\frac 43 \pi R^3 \rho({\rm bulk}) =
4\pi \int_0^{\infty} \rho(r) r^2 dr \,.
\end{equation} 
For its importance in experimental techniques, 
the {\it equivalent rms radius} $Q$ is commonly used. It describes a uniform sharp
distribution with the same rms radius as the given density profile:
\begin{equation} 
\label{q}
\frac 35 \, Q^2= \langle r^2 \rangle \,.
\end{equation} 
Because the neutron skin thickness (\ref{skin}) is defined through rms radii, it can be
expressed easily with $Q$:
\begin{equation}
\label{r0}
\Delta R_{np}=\sqrt{\frac{3}{5}} \left(Q_n-Q_p\right) .
\end{equation}

In uniform, sharp-edge nuclear distributions all of the
above-mentioned definitions coincide, but in realistic leptodermous
density profiles they give distinct values. The relation between the
$C$, $Q$, and $R$ radii can be expressed by expansion formulas in
powers of $b/R$, where $b$ is the so-called surface width of the
density profile. The latter quantity is defined by
\begin{equation} 
\label{b}
b^2= - \frac{1}{\rho(0)} 
\int_0^{\infty} (r-C)^2 \frac{d \rho(r)}{dr} dr \,.
\end{equation} 
To leading order, one has the relationships \cite{has88}
\begin{equation} 
\label{cr} 
C\simeq R\left(1-\frac {b^2}{R^2}\right)
\end{equation} 
and
\begin{equation} 
\label{qr}
Q\simeq R\left(1+\frac 52\frac {b^2}{R^2}\right) .
\end{equation} 
These expansions are useful as long as $b/R$ is small. This condition
is well fulfilled in many nuclei because $b\sim 1$ fm and $R\sim r_0
A^{1/3}$;  therefore, $b^2/R^2$ is typically no larger than
$A^{-2/3}$. Moreover, to consider no further corrective terms in the
above relations of the $C$ and $Q$ radii with $R$ may be quite
accurate for specific shapes. For example, in the case of 2pF
distributions, the first nonvanishing corrections to the terms within
brackets in Eqs.\ (\ref{cr}) and (\ref{qr}) are of order $(b/R)^6$ and
$(b/R)^4$, respectively.

\begin{figure}
\includegraphics[width=0.85\columnwidth,clip=true]
{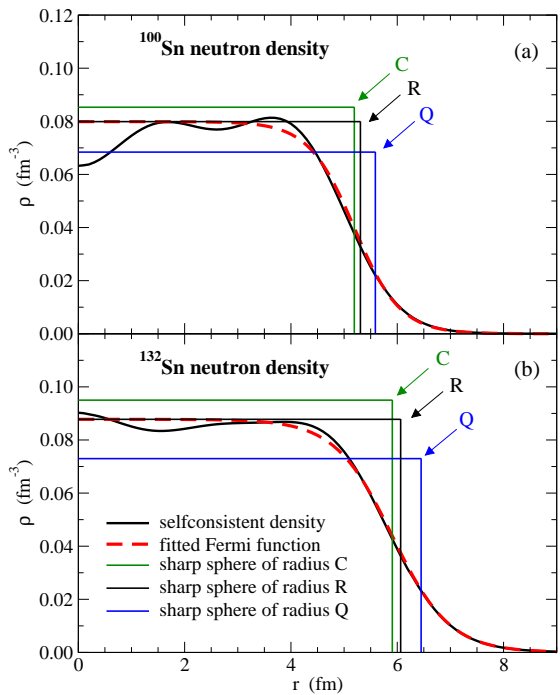}
\caption{\label{radii} 
(Color online)
Comparison of sharp surface density profiles that have the central ($C$),
equivalent sharp ($R$), and equivalent rms ($Q$) radii with the
self-consistent and 2pF profiles corresponding to the neutron density
of (a) $^{100}$Sn and (b) $^{132}$Sn obtained in the RMF theory (the NL3
interaction \cite{lal97} has been used).}
\end{figure}

The multiple definitions of nuclear radii may sometimes cause 
misleading conclusions, especially if nuclear properties sensitive to
the nuclear matter distribution in nuclei are concerned or two different
models are compared. Therefore one has to be careful about the
suitable choice of the radius definition. To compare the
foregoing definitions of radii for heavy nuclei, in Fig.~\ref{radii}
we plot the neutron densities obtained in a self-consistent mean-field
calculation and the fitted 2pF distributions for the neutron-deficient
nucleus $^{100}$Sn and  the neutron-rich nucleus $^{132}$Sn. These
profiles are compared with sharp-edge density distributions having
radii $C$, $R$, and $Q$, calculated from the above expressions with
the 2pF function. The central densities of the sharp
surface spheres are fixed so to fulfill the particle number normalization.

Figure~\ref{radii} illustrates the fact that the central radius $C$
does not allow  the bulk density to be reproduced. A sharp sphere of radius
$C$ overestimates the self-consistent density in the whole nuclear
interior. The equivalent rms radius $Q$ also fails, because it clearly
underestimates the original density in the bulk. Only the equivalent
sharp radius $R$ is able to properly reproduce  the bulk part of the
self-consistent and 2pF density profiles of the nucleus. As discussed
in Ref.\ \cite{has88}, the equivalent sharp radius $R$ is the quantity
of basic geometric importance of the three radii $C$, $R$, and $Q$.
A sharp distribution of radius $R$ has the same volume integral as the
actual density of the finite nucleus and differs from it only in the
surface region.
Therefore, the radius $R$ appears to be  the suitable quantity to be
used to measure the size of the bulk part of the nucleus.

On account of Eq.\ (\ref{r0}) for the neutron skin thickness and 
relationship (\ref{qr}) between $Q$ and $R$, one obtains the expression 
\begin{equation}
\label{r1}
\Delta R_{np} = \sqrt{\frac{3}{5}}\left[(R_n-R_p)
+\frac{5}{2}
\left(\frac {b_n^2}{R_n}-\frac {b_p^2}{R_p}\right) \right] 
\end{equation}
up to terms of order $O(b^4/R^3)$.
Thus, one can make a meaningful distinction between a {\it bulk}
contribution and a {\it surface} ({\it diffuseness}) contribution to
the neutron skin thickness of nuclei as follows:
\begin{equation}
\label{rtot}
\Delta R_{np} = \Delta R_{np}^{\rm bulk} + \Delta R_{np}^{\rm surf} ,
\end{equation}
with
\begin{equation}
\label{rb}
\Delta R_{np}^{\rm bulk} \equiv \sqrt{\frac{3}{5}}\left(R_n-R_p\right) 
\end{equation}
independent of surface properties and
\begin{equation}
\label{rs}
\Delta R_{np}^{\rm surf} \equiv \sqrt{\frac{3}{5}} \, \frac{5}{2}
\left(\frac{b_n^2}{R_n}-\frac{b_p^2}{R_p}\right) .
\end{equation}
The nucleus may develop a neutron skin by separation of the bulk radii
$R$ of neutrons and protons or by modification of the diffuseness $b$
of the neutron and proton surfaces. In the general case a
combination of both effects can be found. To which degree the different patterns
arise in mean-field calculations of finite nuclei is a question we
 address in the following sections.

Experimental \cite{vri87} and theoretical mean-field density
distributions present oscillations in their inner bulk region, which
implies that some suitable average is needed to determine the bulk
density value $\rho({\rm bulk})$ of Eq.\ (\ref{r})  to compute
the equivalent sharp radius $R$. The difficulty may be easily solved
by fitting a Fermi function to the original density, as we have
illustrated in Fig.~\ref{radii}. In this case, the bulk density value
,that is, $\rho_0/[1+ \exp{(-C/a)}]$, is, to excellent accuracy the
$\rho_0$ parameter of the Fermi function. An effect not described by
functions like the 2pF distribution is the occurrence of a
nonsymmetric surface shape, which is possible in real nuclear
densities. However, one plausibly expects that neutron skins of nuclei
are dominated by the difference existing between neutrons and protons
in the location of the surface (i.e., $R$) and/or in the spatial
extent of this surface (i.e., $b$); the difference in the degree of
asymmetry between the shapes of the neutron and proton surfaces is expected 
to be a less important,  higher-order correction.

It may be practical to rewrite expressions (\ref{rb}) and
(\ref{rs}) for the bulk and surface contributions to the neutron skin
thickness directly in terms of the parameters of the 2pF function of
Eq.\ (\ref{2pf}). Using the fact that the 
diffuseness parameter $a$ in a 2pF function is related to the surface
width $b$ by the formula
\begin{equation} 
\label{ba}
b= \frac {\pi}{\sqrt{3}}a 
\end{equation} 
and inverting  relation (\ref{cr}) between $C$ and $R$, one easily
finds to the given order of approximation that Eqs.\ (\ref{rb}) and
(\ref{rs}) become, respectively,
\begin{equation}
\label{r4}
\Delta R_{np}^{\rm bulk}=\sqrt{\frac{3}{5}}\left[(C_n -C_p)
+\frac{\pi^2}{3}
\left(\frac {a_n^2}{C_n}-\frac {a_p^2}{C_p}\right) \right] 
\end{equation}
and
\begin{equation}
\label{rs4}
\Delta R_{np}^{\rm surf} = \sqrt{\frac{3}{5}} \, \frac{5\pi^2}{6}
\left(\frac{a_n^2}{C_n}-\frac{a_p^2}{C_p}\right) .
\end{equation}
We note from these results that
\begin{equation}
\label{r5}
\Delta R_{np}^{\rm bulk} = \sqrt{\frac{3}{5}} (C_n-C_p) 
+ \frac{2}{5} \, \Delta R_{np}^{\rm surf},
\end{equation}
that is, the quantities $C_n-C_p$ (easily obtained from the 2pF
distributions themselves) and $R_n-R_p$ [obtained through Eq.\
(\ref{r})] differ by surface diffuseness terms and should not be mixed.
We have seen in Fig.~\ref{radii} that a sharp sphere having a radius
$C$ significantly distorts the appearance of the actual nuclear
density by overshooting it in the whole bulk region. It is thus
preferable to use $\sqrt{3/5} (R_n-R_p)$ rather than $\sqrt{3/5} (C_n-C_p)$
as a measure of the bulk contribution to the neutron skin thickness.


\section{Decomposition of neutron skin thickness in selected nuclei
- mean-field results}

\subsection{Nuclei of the experiments in antiprotonic atoms}\label{IVA} 

\begin{figure}
\includegraphics[width=0.95\columnwidth,clip=true]
{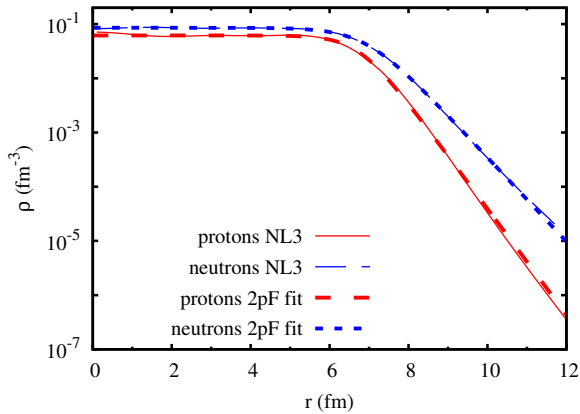}
\caption{\label{FIGURE2} (Color online)
The proton and neutron densities of
$^{208}$Pb calculated with the NL3 parameter set as a function of
the distance from the center of this nucleus in logarithmic scale. The
2pF densities fitted to the NL3 results are also shown.}
\end{figure}

To get information about the ``bulk'' or ``surface''
character of the thickness of neutron skins in theoretical mean-field
calculations, we parametrize the self-consistently calculated proton
and neutron densities with 2pF distributions. 
This procedure can be applied very suitably for many heavy nuclei and gives a clear
distinction between bulk and surface properties of nuclei. 
However there is no universal method to do this parametrization.
A popular prescription is to use a $\chi^2$ minimization of the 
differences
between the density to be reproduced and the 2pF profile, or of the
differences between their logarithms. These methods may somewhat
depend on conditions given during minimization (number of mesh points,
limits, etc.). We prefer to extract the parameters of the 2pF
profiles by imposing that they reproduce the same quadratic $\langle
r^2 \rangle$ and quartic $\langle r^4 \rangle$ moments of the
self-consistent mean-field densities. These two conditions, together
with the normalization to the proton and neutron numbers, allow us to
determine in a unique way the equivalent 2pF densities. This
method can be applied to any density distribution. Its focus is on
a good reproduction of the 
surface region of the original density because the local 
distributions of the quantities $r^2 \rho(r)$ and $r^4 \rho(r)$ 
are peaked at the periphery of the nucleus.
As an example of the results of the present determination of the 2pF
profiles, in Fig.~\ref{FIGURE2} we display in logarithmic scale the
self-consistent neutron and proton densities for $^{208}$Pb together
with their equivalent 2pF distributions calculated with the NL3
interaction. It can be seen that there is  overall good
agreement in the central and surface regions of the nucleus. In
particular, the 2pF densities reproduce  the mean-field
densities well at the distances that are relevant for antiproton
annihilation.

In this section, we would like to get some insight about the ``bulk'' or
``surface'' character of the neutron skin predicted by the mean-field
approach in the nuclei for which the neutron skin thickness values
were extracted in Refs.\ \cite{trz01,jas04} from the measurements in
antiprotonic atoms. These nuclei range from $^{40}$Ca to $^{238}$U and
all lie along the valley of stability. We carry out the
calculations with the aforementioned nonrelativistic interactions D1S
\cite{ber91} and SLy4 \cite{cha98} plus the relativistic interactions
NL3 \cite{lal97} and FSUGold \cite{FSUG}, as representative examples
of successful nuclear mean-field models. 
We impose spherical symmetry in all nuclei described in this article.
The effect of deformation on the neutron skin thickness was
discussed elsewhere \cite{war98}.
The two nonrelativistic
forces have a soft symmetry energy \cite{ste05a,li08}. 
On the contrary, the covariant NL3 parameter set has
a stiff symmetry energy, which is
usual in the relativistic models \cite{ste05a,li08}.
Note that the notation
soft or stiff refers to whether the symmetry energy of the model
increases slowly or rapidly as a function of the nuclear density. 
The covariant parameter set FSUGold was devised to have a softer
density
dependence of the symmetry energy \cite{FSUG} than the typical
relativistic models. Thus, FSUGold is, in this aspect, closer to the
nonrelativistic models than NL3.

\begin{figure}
\includegraphics[width=0.95\columnwidth,clip=true]{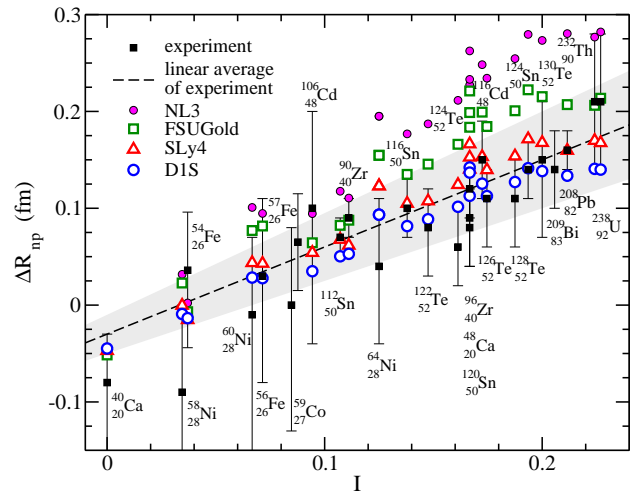}
\caption{\label{FIGURE3} (Color online)
The results of the covariant NL3 and FSUGold parameter sets and of the
nonrelativistic Skyrme SLy4 and Gogny D1S forces  compared with the
experimental neutron skin values $\Delta R_{np}$ deduced from
antiprotonic atoms (solid squares with errorbars) and their linear
average $\Delta R_{np}= (0.90 \pm 0.15) I + (-0.03 \pm 0.02)$ fm
(shaded region) \cite{trz01,jas04}.}
\end{figure}

In Fig.~\ref{FIGURE3} we display the experimental data with 
error bars determined from the antiprotonic atoms. There is a
relatively clear correlation between the experimental value of the
neutron skin thickness of these 26 stable nuclei and the overall
relative neutron excess $I=(N-Z)/A$ of the nucleus. This trend has
been fitted by the linear relationship $\Delta R_{np}= (0.90 \pm 0.15)
I + (-0.03 \pm 0.02)$ fm with a $\chi^2$ factor of 0.5
\cite{trz01,jas04}. In the same figure we plot the theoretical $\Delta
R_{np}$ value calculated according to Eq.\ (\ref{skin}) using the
mean-field densities of the indicated effective nuclear models for 
the 23 even-even nuclei that exist in the
experimental data set. It is obvious that the theoretical models make
largely different predictions for the neutron skin thickness. This is
especially visible at large values of~$I$. It is seen that the models
that have a softer symmetry energy give smaller $\Delta R_{np}$
values, whereas the models with a stiffer symmetry energy give larger
$\Delta R_{np}$ values \cite{cen09,cen09b}. Note that as soon as
$I\ne0$, even when it is small, discrepancies  arise  among the
$\Delta R_{np}$ values of
the models. In contrast,  all of the considered
interactions make an almost identical prediction of
$\Delta R_{np}\approx-0.05$~fm for the $^{40}$Ca nucleus.

\begin{figure*}[bth]
\includegraphics[width=120mm]{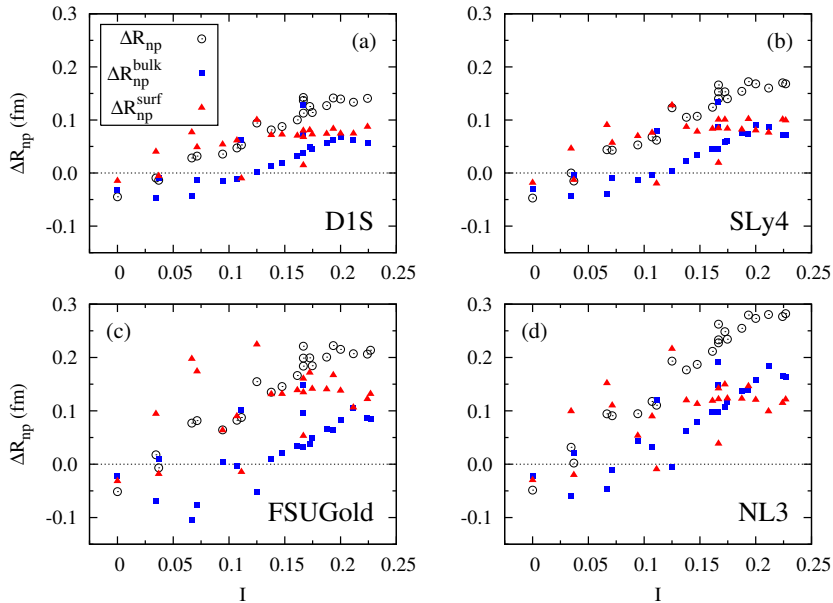}
\caption{\label{FIGURE4} (Color online)
The neutron skin thickness $\Delta R_{np}$ as
well as its bulk $\Delta R_{np}^{\rm bulk}$ and surface $\Delta
R_{np}^{\rm surf}$ parts [given by Eqs.\ (\ref{rb}) and (\ref{rs}),
 respectively] for the nuclei presented in
Fig.~\ref{FIGURE3}, calculated in various models.}
\end{figure*}
 

One observes in Fig.~\ref{FIGURE3} that, similarly to the situation
in the experimental data set, the theoretical neutron skin values
computed with each nuclear interaction show an average linear behavior
as a function of the neutron excess $I$. Actually, the linear
correlation factor of $\Delta R_{np}$ with $I$ in the present models
is considerably high (between 0.95 and 0.97). The fit of the
neutron skin values calculated with the Skyrme SLy4 force yields
$\Delta R_{np}= 1.01 I -0.035$ fm, whereas $\Delta R_{np}= 0.88 I
-0.04$ fm is obtained in the case of the Gogny D1S interaction. 
We find a linear fit of 
$\Delta R_{np}= 1.20 I -0.03$ fm using the values of the neutron skin
thickness computed with the relativistic FSUGold model. If we
consider the relativistic NL3 parameter set, which has a stiffer
symmetry energy than the other models, we find the linear fit $\Delta
R_{np}= 1.50 I -0.03$ fm. Compared with the slope of 0.90 for the fit of
the experimental data \cite{trz01,jas04}, the slopes in forces like D1S
and SLy4 is much closer to it, whereas the agreement deteriorates as
the model has a stiffer symmetry energy. Thus, one can conclude that
the comparison of the slopes of $\Delta R_{np}$ with $I$ between
theory and the antiprotonic measurements for nuclei across the mass
table favors the models that have a soft symmetry energy.

In Fig.~\ref{FIGURE4} we display the bulk and surface contributions to
the theoretical neutron skin thickness that are computed by applying
Eqs.\ (\ref{rb}) and (\ref{rs}), respectively, against the neutron
excess $I$. These  values are obtained from the 2pF distributions
associated with the mean-field neutron and proton densities calculated
with the D1S force as well as with the Skyrme SLy4
force and the two RMF parameter sets FSUGold and NL3. 
The numerical calculations show that the value of $\Delta R_{np}^{\rm bulk} +
\Delta R_{np}^{\rm surf}$ obtained through Eqs.\ (\ref{r4}) and
(\ref{rs4}) can be slightly less accurate in some nuclei  [compared to
the exact value of Eq.\ (\ref{skin})] than the result obtained through
Eqs.\ (\ref{rb}) and (\ref{rs}). The small differences, whenever they
arise, are due almost entirely to the replacement of $a_q^2/R_q$ by
$a_q^2/C_q$ ($q=n,p$) in Eqs.\ (\ref{r4}) and (\ref{rs4}). Of course,
Eqs.\ (\ref{r4}) and (\ref{rs4}) are quite practical because they do not
require the additional calculation of the $R_q$ values and can be
applied straightforwardly using the $C_q$ and $a_q$ parameters of the
2pF distributions themselves.

The four panels presented in Fig.~\ref{FIGURE4}, 
as a consequence of the differences between models, 
predict a slightly different splitting of the neutron skin thickness
into their bulk and surface contributions 
for each considered nucleus. Therefore, the values of the
bulk and surface contributions to $\Delta R_{np}$ are to some extent
model dependent. 
The total neutron skin thickness roughly follows an increasing tendency
with the relative neutron excess $I$, as discussed previously.
This tendency is also seen in the bulk contribution $\Delta R_{np}^{\rm bulk}$, 
especially in the nonrelativistic interactions. However 
the surface contribution does not clearly follow  the same trend and shows a 
less definite behavior as a function of $I$.

From Fig.~\ref{FIGURE4} we see that for relatively neutron-rich nuclei
($I\gtrsim 0.15$), the surface part generally contributes 50\% or more to
the total neutron skin thickness. If the nuclear model has a stiff
symmetry energy and $I$ is large, the bulk part of $\Delta R_{np}$
may become larger than the surface part, as  can be seen 
in the NL3 panel of Fig.~\ref{FIGURE4} at $I>0.2$. More
symmetric nuclei ($I\lesssim 0.15$) do not present a definite tendency.
The theoretical 
calculations show that in these nuclei the  
sharp radius is larger for protons than for neutrons ($R_p>R_n$) while
the surface width is larger for neutrons than for protons ($b_n>b_p$).
Consequently, the bulk contribution to the neutron skin thickness
becomes negative, as it can be seen in Fig.~\ref{FIGURE4}, and the
relatively small value of the neutron skin thickness is basically because
of a strong cancellation between the bulk and surface parts. In
the lightest nuclei, both contributions are negative and produce a
``proton skin" rather than a neutron skin.


\subsection{Medium and heavy mass isotopic 
chains}


The set of nuclei chosen in Sec. IV A have given us some 
insight into the bulk and surface contributions to the neutron skin
thickness in stable isotopes. To investigate in a systematic way how the
bulk and surface contributions  evolve with the neutron number we study
the chains of even-even Sn and Pb isotopes from the proton to the neutron drip
line.

\begin{figure}
\includegraphics[width=70mm]{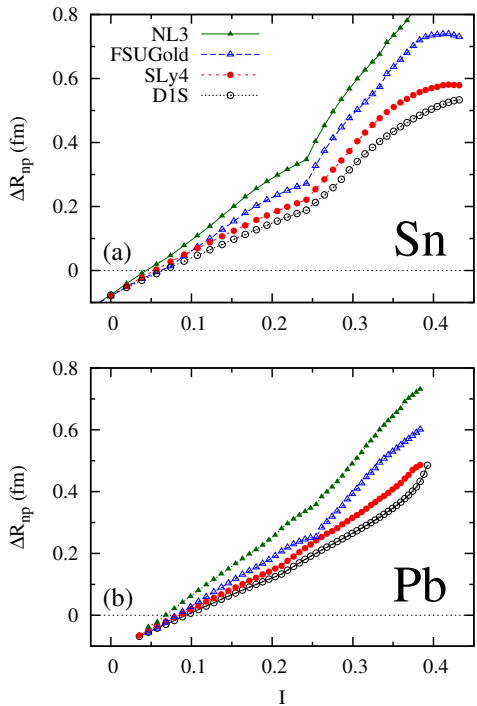}
\caption{\label{FIGURE5} (Color online)
The neutron skin $\Delta R_{np}$ for (a) Sn 
and (b) Pb isotopes calculated with several mean-field models.}
\end{figure} 

\begin{figure}
\includegraphics[width=0.95\columnwidth]{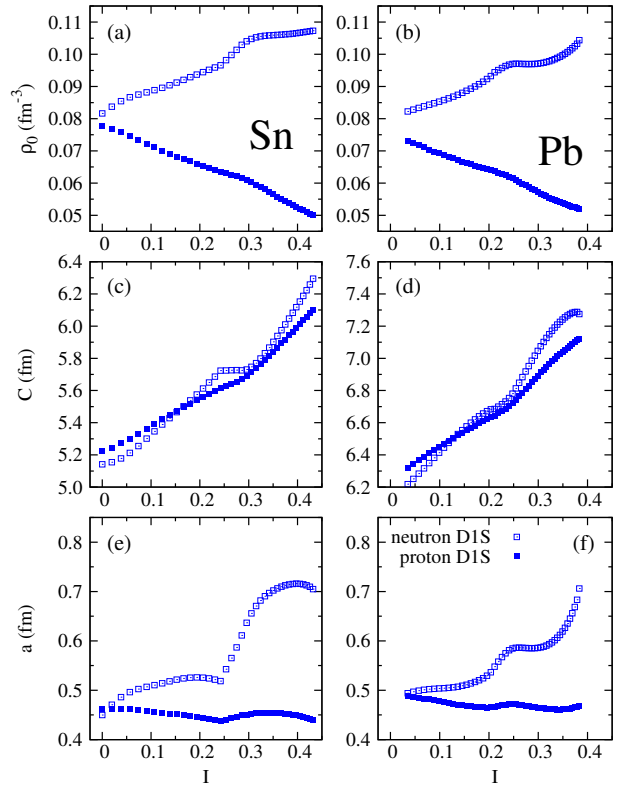}
\caption{\label{FIGURE7} (Color online)
Proton and neutron 2pF parameters (a) $\rho_0$ for Sn, (b) $\rho_0$ for Pb,
 (c) $C$ for Sn, (d) $C$ for Pb, (e) $a$ for Sn,  and  (f) $a$ Pb  isotopes. 2pF parameters are
fitted to the Gogny D1S density distributions.
}
\end{figure}

\begin{figure}
\includegraphics[width=0.95\columnwidth]{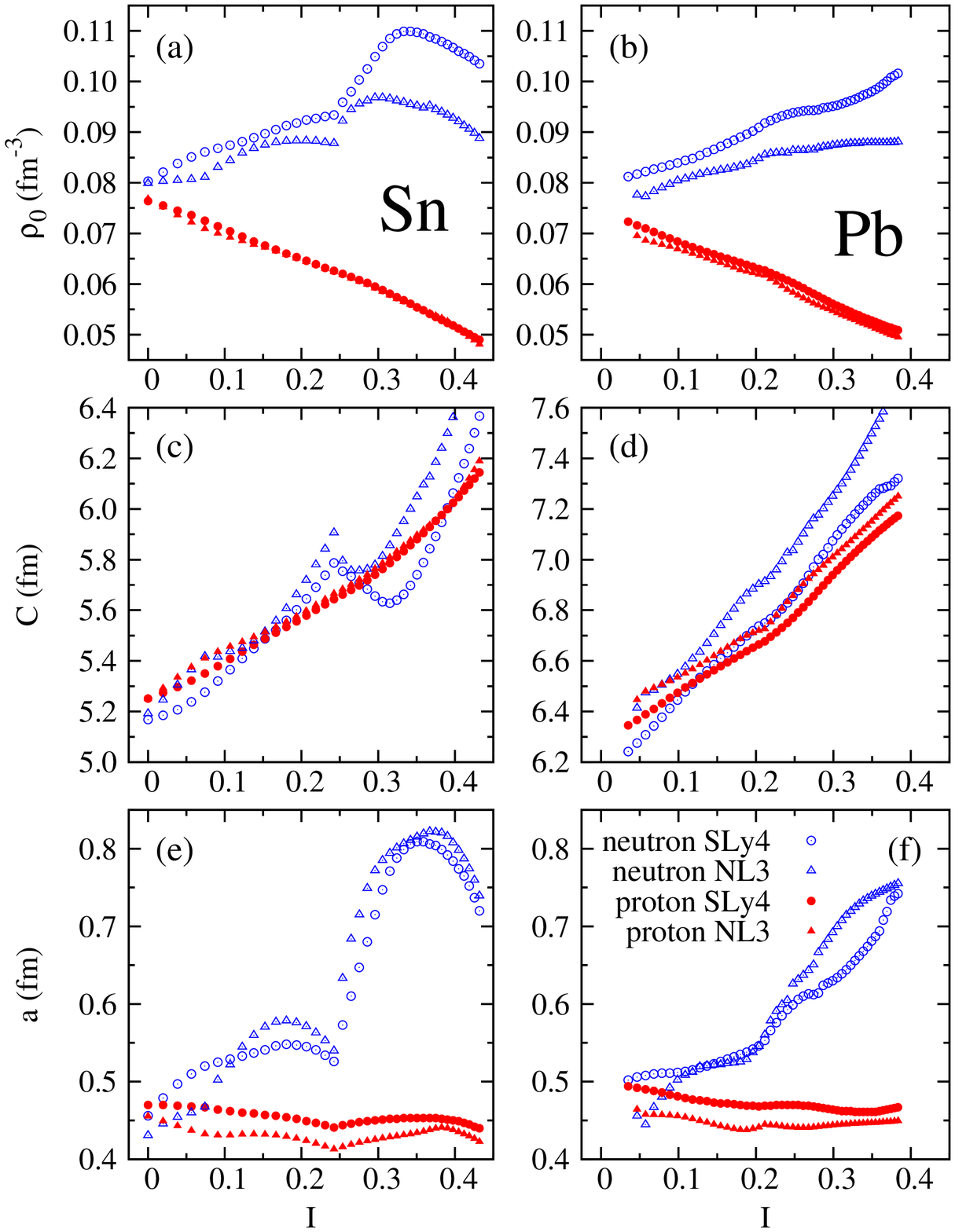}
\caption{\label{FIGURE8} (Color online)
The same as in Fig.~\ref{FIGURE7} but for the Skyrme  SLy4 
force and relativistic NL3 parameter set.
}
\end{figure}

First,  Fig.~\ref{FIGURE5} displays the neutron skin thickness
$\Delta R_{np}$ along the Sn and Pb isotopic chains computed in the
mean-field  approximation with the Gogny D1S and Skyrme SLy4 forces and
using the  relativistic mean-field models NL3 and FSUGold. For small and
moderate values of the relative neutron excess ($I\lesssim 0.2$), the
neutron skin thickness grows almost linearly with $I$ in each isotopic
chain, as in the case of the stable nuclei analyzed in the previous
section. However, $\Delta R_{np}$ shows a rather pronounced kink at
$I\approx 0.20-0.25$.  Beyond this value, it increases again almost
linearly as a function of the relative  neutron excess, but with a
larger slope. Finally, a new departure from the linear behavior of
$\Delta R_{np}$ as a function of $I$ can be observed when the isotopes
are on the edge of the neutron drip line.  The kinks and changes of
slope in the neutron skin thickness as a function of  the relative
neutron excess are clearly connected with the doubly magic  nuclei 
$^{132}$Sn ($I\approx0.24$) and $^{208}$Pb ($I\approx0.21$) in the
stable  nuclei region and with another two doubly magic nuclei, namely
$^{176}$Sn ($I\approx0.43$) and $^{266}$Pb ($I\approx0.38$), near the
neutron drip line.  Therefore, it is obvious that these changes of slope
are produced by quantal effects, which modify the linear trend of  
$\Delta R_{np}$ as a function of $I$ in a non-negligible way. The average
slope of $\Delta R_{np}$ with $I$ for various forces is clearly
different.  As  has been shown in Refs.\cite{cen09,cen09b}, for each
nuclear model, $\Delta R_{np}$ is strongly correlated with the slope of
the symmetry energy with respect to the density computed at saturation.

\begin{figure*}
\includegraphics[width=120mm]{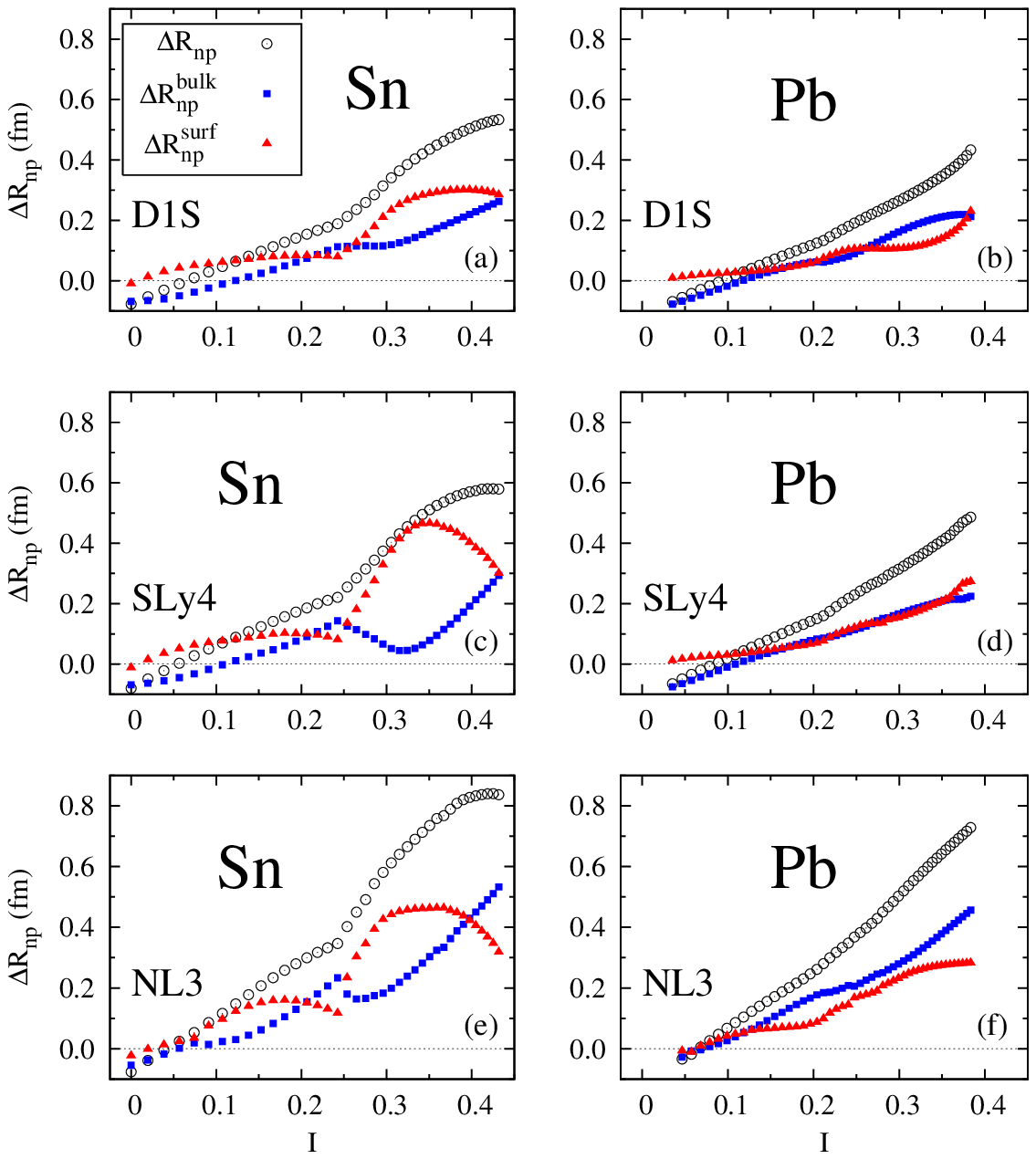}
\caption{\label{FIGURE8a} (Color online)
The same as in Fig.~\ref{FIGURE4} but for Sn and Pb
 isotopes calculated with the Gogny D1S force, Skyrme  SLy4 force, and relativistic NL3 parameter set.}
\end{figure*}

As in Sec. IV A, we fit the mean-field proton
and neutron densities by 2pF distributions to investigate the bulk and
surface contributions to the neutron skin. To analyze the changes that occur
in $\Delta R_{np}$ first we look at the parameters that characterize 
the 2pF distributions. In Fig.~\ref{FIGURE7} we
display the central density $\rho_0$, the half-density radius $C$ and
the diffuseness $a$, which are obtained from the D1S densities 
along the Sn and Pb isotopic chains.
The 2pF parameters show an overall smooth behavior as a function of
the relative neutron excess $I$ but with local modulations
near the shell closures. The central density $\rho_0$ grows (for
neutrons) or declines (for protons) almost linearly as a function of
$I$ in both isotopic chains. It is just a simple consequence of the
increasing asymmetry along the isotopic chains.
The central radii $C_n$ and $C_p$ both show  a global increasing
tendency with the neutron number. One can notice that $C_n$ grows
faster with $I$ than $C_p$. However, the $a_n$ and $a_p$
diffuseness parameters behave in a completely different way. Whereas
$a_n$  grows on average with increasing neutron excess,
$a_p$  remains roughly constant with $I$ in the two
isotopic chains. On top of these general trends we see that the 2pF
parameters associated with the neutron densities show kinks around the
shell closures at $N=82$ in Sn and $N=126$ in Pb (that is, in the
region $I\simeq0.20-0.25$), as well as changes of slope near the
neutron drip lines. One observes that some signature of the neutron
shell effect also appears  in the 2pF parameters of the proton
densities around magic neutron numbers.

Figure~\ref{FIGURE8} shows the evolution with $I$ of the 2pF
parameters corresponding to the quantal densities computed with the
SLy4 Skyrme force and with the relativistic mean-field NL3 parameter
set. In the two models the global trends are similar to the ones discussed before
for 2pF parameters of D1S distribution. 
However, they show some differences that can be
attributed in part to the different properties of the isovector
channel of the interaction.

In  Figs.~\ref{FIGURE8a}(a) and \ref{FIGURE8a}(b) we display
the bulk and the surface contributions to the nuclear skin thickness
calculated with the D1S force along the Sn and Pb isotopic chains.
These contributions show for Sn isotopes two well-defined
regions as a function of $I$. One of them covers the neutron major shell between
$^{100}$Sn and $^{132}$Sn, i.e. $0\leq I\lesssim0.25$, and the other
region corresponds to the next major shell between $^{132}$Sn and
$^{176}$Sn in the range $0.25\lesssim I\lesssim0.43$. 
For nearly symmetric Sn isotopes close to $^{100}$Sn, which are
neutron deficient, $C_p$ is larger than $C_n$  (see Fig.~\ref{FIGURE7}) 
and, consequently, the bulk part of the neutron
skin is negative or at most it takes very small positive values. For
these values of $I$, the surface contribution is positive and
relatively small, reducing the opposite effect of ``proton skin'' due
to negative values of the bulk part. 
Looking at more asymmetric isotopes, one can see that,
in the magic nuclei $^{132}$Sn and $^{176}$Sn, the bulk and the surface
contributions are roughly equal. This implies a rather compact neutron
density distribution with a relatively stiff surface as a consequence
of the kinks exhibited by the neutron diffuseness parameter $a_n$
around the neutron magic numbers, which can be seen in Fig.~\ref{FIGURE7}.
In the regions between magic numbers, the surface contribution to the
neutron skin thickness is larger than the bulk contribution. The splitting
between the surface and bulk contributions reaches its maximum value
roughly at midshell.
This behavior of the bulk and surface contributions to the neutron
skin thickness points out that Sn isotopes in the middle of major
shells develop a larger surface region in the density distribution
than the magic ones. In other words, these
isotopes with neutron number in between  magic values are more of
``halo'' type than the limiting magic nuclei which show a mixed
character between ``halo''  and ``neutron skin.'' 

The bulk and the surface contributions to the neutron skin of the Pb 
isotopes calculated with the D1S force show a similar 
behavior to the case of the Sn isotopes analyzed before. However, for
this heavy isotopic chain, the bulk part gives a more important
contribution to the total neutron skin for nuclei in between  the
magic $^{208}$Pb and $^{266}$Pb nuclei.

The discussed differences in the behavior of the neutron skin and its bulk
and surface
contributions along the neutron-rich Sn and Pb isotopes can be qualitatively
understood as follows.
Within a major shell, the rms radii of the different single-particle
orbits are spread around their average value. The rms radius of orbits
with low angular momentum $l$ are larger than the average, while the
rms radius of orbits with high $l$ are slightly smaller but close to
the average in the shell.
Consequently, the outermost region  of the density is basically provided by the
orbits with low $l$ in the last populated major shell. On the contrary, orbits 
with high $l$ in this major shell have their most important contribution at 
shorter distances from the center of the
nucleus, increasing more the bulk than the surface part of the nuclear
density. Hence, the filling order of the last single-particle orbits is crucial 
to 
determine if the growing of the neutron radius, and consequently the neutron 
skin, is due to an increase of the surface or the bulk of the density.
In the case of neutron-rich isotopes of Sn above $N=82$, the first filled orbits 
are $2f_{7/2}$, $3p_{3/2}$ and $3p_{1/2}$. This ordering produces 
an important 
enhancement of the nuclear surface, which can be appreciated from Figs.\ \ref{FIGURE7}  and \ref{FIGURE8a}.
Once midshell is filled, the $1h_{9/2}$ and $1i_{13/2}$ orbits start to be appreciably 
populated, increasing the bulk more than the surface of the densities and 
hence of the neutron 
skin, as  can be seen in the aforementioned figures. The situation in Pb isotopes 
is just the contrary; above $N=126$, the first occupied levels are $2g_{9/2}$ and 
$1j_{15/2}$,  which increase the bulk more than the surface. Only near the drip 
line, $N=184$, are  the low-momentum orbits $3d_{5/2}$, $4s_{1/2}$, and $3d_{3/2}$
 relevant, increasing the surface and quenching the bulk contributions 
to the density and consequently of the neutron skin. This behavior in Pb isotopes 
can also be seen in Figs.\ \ref{FIGURE7}  and \ref{FIGURE8a}.

In Figs.~\ref{FIGURE8a}(c)-\ref{FIGURE8a}(f) we display the results 
calculated with the Skyrme SLy4 force and the NL3 parameter set. 
The bulk and surface
contributions qualitatively behave as the ones computed with the D1S
force discussed previously. Of course, some differences are
found in comparing the results obtained with the different models because of
their different isovector properties. 
In models with a stiff density dependence of the symmetry energy
(for instance, NL3), the bulk contribution to the neutron skin is more
important than in models with a soft symmetry energy, as in the case
of the SLy4 and D1S forces. This tendency can be especially noted in
the Pb isotonic chain.


\section{Conclusions}

Ground-state properties of stable nuclei, such as charge radii and
binding energies, can be reproduced fairly well by using mean-field
models with effective interactions such as the Skyrme or Gogny forces
and with relativistic Lagrangians. Although the isoscalar part of
these models is well constrained, their isovector properties are much
less determined and the predictions in this sector can differ
considerably, even for stable nuclei. A typical example is the neutron
skin thickness of nuclei. For this observable theoretical predictions
for $^{208}$Pb using nonrelativistic forces give a value around 0.15
fm while relativistic mean-field parametrizations predict values that
are almost twice as large.

The neutron skin thickness of a set of 26 nuclei, distributed over the
whole periodic table, has been obtained from  the analysis of
experiments with antiprotonic atoms \cite{trz01,jas04} combined with the
charge radii obtained from electron scattering data. One important
result of these experiments with antiprotonic atoms is that there
exists  a rather clear linear correlation between the neutron skin
thickness and the overall neutron excess $I$. Theoretical mean-field
calculations of the neutron skin also show this tendency even more clearly.
 It is found that the relativistic parametrizations
systematically  predict larger neutron skins than the ones computed
with  nonrelativistic interactions. This is because the 
symmetry energy is stiffer in the relativistic models than in the
nonrelativistic models.

To analyze the experimental data of antiprotonic atoms an ansatz of the
nuclear densities is needed.  The two-parameter Fermi distributions have
been used often to this end. It is found that the  experimental data can
be reproduced by a variety of these Fermi distributions with different
values of   $C_n-C_p$ and $a_n-a_p$. The experimental values of  the
halo factor in $^{208}$Pb are well reproduced by distributions with 
$C_n\approx C_p$ (halo model) and by Fermi densities with both
halo and neutron skin ($a_n\approx a_p$)  contributions. This
latter scenario is also well predicted  by the nonrelativistic
mean-field densities obtained with the D1S  and SLy4 forces.

We  have also parametrized the mean-field densities via two-parameter
Fermi distributions. We do this by imposing that both mean-field and
parametrized densities give the same quadratic and quatric moments.
This parametrization of the mean-field densities also allows 
 the neutron skin thickness to be split
easily into two contributions, namely the bulk
part and the surface part.
It is found that the mean-field neutron skins computed
in nuclei with $I>0.1$ can be shared between non-negligible
surface and bulk parts. This applies both for stable nuclei investigated
in antiprotonic experiments and for drip line isotopes in all the theoretical models
considered in this work.

To analyze the neutron skin in neutron-rich nuclei, we have
theoretically studied its variation along the Sn and Pb isotopic chains
up to the neutron drip line using selected mean-field models. As expected,
$\Delta R_{np}$ shows  generally linear growing trend with  $I$. 
However, shell effects, which are always present in
mean-field calculations, produce noticeable departures  from this linear
dependence in nuclei with large neutron excesses.

Regarding the bulk and surface contributions to the neutron skin
thickness in Sn isotopes, it can be seen that the considered mean-field
models point toward more of a surface character in stable nuclei. This
effect is reinforced in the neutron-rich region. In the case of the Pb
isotopic chain, bulk and surface contributions have similar values in
stable isotopes, whereas the bulk part is larger than the surface part
in the more neutron-rich region of this isotopic chain.

\begin{acknowledgments}

This work was partially supported by the Spanish
Consolider-Ingenio 2010 Programme Centro Nacional de
F´ısica de Part´ıculas, Astropart´ıculas y Nuclear CSD2007-
00042 and by Grant Nos. FIS2005-03142 and FIS2008-01661
from Ministerio de Educaci´on y Ciencia (Spain) and Fondo
Europeo de Desarrollo Regional, Grant No. 2009SGR-1289
from Generalitat de Catalunya (Spain), and Grant No. N N202
231137 from Ministerstwo Nauki i Szkolnictwa Wy˙zszego
(Poland). X.R. also acknowledges Grant No. AP2005-4751
from MEC (Spain).
\end{acknowledgments}

\appendix
\section{}


In two-parameter Fermi (2pF) density distributions 
\begin{equation} 
\label{EQUATION_A0}
\rho_q(r) = \frac{\rho_{0q}}{1+ \exp{[(r-C_q)/a_q]}} \,,
\end{equation} 
the number of
particles and the mean square radii can be approximated, respectively,
by the relationships
\begin{equation} 
\label{EQUATION_A1}
N_q=\frac 43 \pi C_q^3 \rho_{0q} 
     \left( 1 + \frac{\pi^2 a_q^2}{C_q^2} \right)
\end{equation} 
and
\begin{equation} 
\label{EQUATION_A2}
\langle r^2\rangle_q=\frac 35 C_q^2 
 \left( 1+\frac 73 \frac{\pi^2 a_q^2}{C_q^2} \right) ,
\end{equation} 
%
%
%
where, $q=n,p,c$ denotes the neutron, proton, and charge distributions,
respectively.
The result of Eq. (\ref{EQUATION_A2}) can be easily derived by recalling
Eqs.\ (\ref{q}), (\ref{cr}), (\ref{qr}), and (\ref{ba}) given in 
Sec.~\ref{discerning}. 

From the relation existing between the charge and proton rms radii
\begin{equation} 
\label{EQUATION_A4}
\langle r^2\rangle_c=\langle r^2\rangle_p+0.64 \;\mathrm{fm}^2 ,
\end{equation} 
together with normalization condition (\ref{EQUATION_A1}), one can
derive the parameters $a_p$ and $C_p$ of the 2pF point proton
distribution if the parameters $a_c$ and $C_c$ of the experimental
charge distribution are known (assuming that the central density
is the same for charge and protons). The result for $a_p$ is
\begin{equation} 
\label{EQUATION_A5}
a_p=\frac{C_p}{\pi}\sqrt{\frac{3Z}{4\pi C_c^3 \rho_{0c}}-1}
\end{equation} 
and $C_p$ is obtained as
\begin{equation} 
\label{EQUATION_A6}
C_p=S_1+S_2 \,,
\end{equation} 
where $S_1$ and $S_2$ are given by the equations
\begin{equation} 
\label{EQUATION_A7}
S_1^3= T_1 + \sqrt{T_1^2+T_2^3} \,,
\end{equation} 
\begin{equation} 
\label{EQUATION_A8}
S_2^3= T_1 - \sqrt{T_1^2+T_2^3} \,,
\end{equation} 
with
\begin{equation} 
\label{EQUATION_A9}
T_1= \frac{21Z}{32\pi \rho_{0c}} \,,
\qquad
T_2= \frac5{12} 
\left( \langle r^2\rangle_c - 0.64 \;\mathrm{fm}^2 \right) . 
\end{equation} 

Once $a_p$, $C_p$, and the rms radius of the point proton density are
available, the following relationship between th eparameters $a_n$ and $C_n$
 of the neutron distribution can be applied:
\begin{equation} 
\label{EQUATION_A11}
a_n^2=\frac5{7\pi^2}
\left( \Delta R_{np}+ \langle r^2\rangle_p^{1/2}\right)^2
-\frac 37 \frac {C_n^2}{\pi^2} \,.
\end{equation} 
This expression is obtained by inserting the neutron skin thickness
$\Delta R_{np}= \langle r^2 \rangle_n^{1/2} - \langle r^2
\rangle_p^{1/2}$ of the nucleus in Eq.\ (\ref{EQUATION_A2}) for $q=n$.
Therefore, for the same $\langle r^2\rangle_p^{1/2}$ and $\Delta
R_{np}$ values, one has a degenerate family of 2pF neutron densities
depending on one parameter, which can be taken to be either $C_n-C_p$
or $a_n-a_p$ (recall that $C_p$ and $a_p$ are known). Alternatively,
if the values of $\langle r^2\rangle_p^{1/2}$ and $C_n-C_p$ are 
given,
one obtains a family of $\Delta R_{np}$ values depending on the
parameter $a_n-a_p$. As mentioned in the main text, the experiments in
antiprotonic atoms were shown to preferentially support the situation
of 2pF density distributions having $C_n-C_p\approx0$ 
\cite{trz01,jas04}.



\end{document}